%% file: eps03lhclc.tex
\documentclass[epj]{svjour}
\usepackage{graphics}
\input defs
\begin{document}
\title{Interplay between the LHC and a Linear Collider in Searches for
New Physics}
\author{Georg~Weiglein 
}                     
%
%
\institute{Institute for Particle Physics Phenomenology, University of
Durham, Durham DH1~3LE, UK}
\date{Received: date / Revised version: date}
%
\abstract{
The LHC / LC Study Group investigates how analyses at the LHC could
profit from results obtained at a future Linear Collider and vice versa,
leading to mutual benefits for the physics program at both machines.
Some examples of results obtained within this working group so far 
concerning searches for new physics are briefly summarised.
} 
\authorrunning{Georg Weiglein}
\maketitle
\section{Introduction}
\label{intro}

Physics at the LHC and a future Linear Collider (LC) will be
complementary in many respects, similarly to the situation at previous
generations of hadron and lepton colliders. The LHC has a large mass
reach for direct discoveries, which extends
up to typically $\sim$6--7~TeV for singly-produced particles. The
hadronic environment at the LHC, on the other hand, 
will be experimentally challenging. 
Owing to the composite nature of the colliding 
protons, LHC physics suffers from the presence of an underlying event
being related to the interaction of the spectator partons.  
Kinematic reconstructions are normally restricted to the transverse
direction. Since the initial-state particles carry colour charge, QCD
cross sections at the LHC are huge, giving rise to backgrounds which are
many orders of magnitude larger than the typical signal processes being 
mostly of electroweak nature. 

The envisaged LC in the energy range of
$\sim$0.5--1~TeV provides a much cleaner experimental environment being
well suited for high-precision physics. It has a well-defined initial
state which can be prepared to enhance or suppress certain processes
with the help of beam polarisation. The better knowledge of the momenta
of the interacting particles gives rise to kinematic constraints which
allow to reconstruct the final state in detail. The signal--to--background
ratios at the LC are in general much better than at the LHC. Direct
discoveries at the LC are possible up to the kinematic limit of the
available energy, while in many cases the indirect sensitivity to effects of
new physics via precision measurements is significantly larger.

While qualitatively the complementarity between LHC and LC is obvious, 
more quantitative analyses of the possible interplay between LHC and LC 
were lacking until recently. In order to investigate this issue, the
so-called LHC / LC Study Group~\cite{lhclc}
has formed as a collaborative effort of
the hadron collider and linear collider communities. This world-wide 
working group investigates in particular how analyses carried
out at the LHC could profit from results obtained at a LC and vice
versa, leading to mutual benefits for the physics program at both machines.

While the LHC is scheduled
to take first data in 2007, the LC could go into operation at about the
middle of the next decade. This would guarantee a substantial period of
overlapping running of both machines, since it seems reasonable to
expect that the LHC (including upgrades) will run for about 20 years.
During simultaneous running of both machines there is obviously the highest
flexibility for adapting analyses carried out at one machine according
to the results obtained at the other machine. The LC results could in
this context also give essential input for choosing suitable upgrade
options for the LHC.

The results obtained so far in the framework of the LHC / LC Study
Group, based on the work of more than 100 contributing authors,
will be documented in a working group report~\cite{lhclcrep}
which is currently being
compiled. Topics under study comprise the physics of weak and strong
electroweak symmetry breaking, electroweak and
QCD precision physics, the phenomenology of Supersymmetric models, 
new gauge theories and models with extra dimensions.
In this talk some examples of the work carried out in the 
LHC / LC Study Group concerning searches for physics beyond the Standard
Model (SM) are briefly summarised.


\section{Some examples of LHC / LC interplay in new physics searches}

\subsection{Determination of SUSY parameters at LHC / LC}

The search for Supersymmetric particles is an example making the need
for an interplay between LHC and LC particularly apparent. The production of
Supersymmetric particles at the LHC will be dominated by the production
of coloured particles, i.e.\ gluinos and squarks.
Searches for the signature of jets and missing energy at the LHC will cover
gluino and squark masses of up to 2--3~TeV~\cite{lhcexp}. The main
handle to detect uncoloured particles will be
from cascade decays of heavy gluinos and squarks, since in most 
scenarios of Supersymmetry (SUSY) the uncoloured particles are
generically lighter
than the coloured ones, e.g.\
$\tilde g \to \bar q \tilde q \to \bar q q \tilde\chi_2^0 \to
\bar q q \tilde\tau \tau \to \bar q q \tau\tau \tilde\chi_1^0$,
where $\tilde\chi_1^0$ is assumed to be
the lightest Supersymmetric particle (LSP).
Thus, the production of Supersymmetric particles at the LHC 
will normally lead to
complicated final states, and in fact the main background for measuring 
SUSY processes at the LHC will be SUSY itself. 

The LC, on the other hand, has good prospects for the production of
uncoloured particles. The clean signatures and small backgrounds at the
LC as well as the possibility to adjust the energy of the collider to
the thresholds at which SUSY particles are produced will allow a precise
determination of the mass and spin of Supersymmetric particles and of
mixing angles and complex phases~\cite{lctdrs}. 

In order to establish SUSY experimentally, it will be indispensable to
verify its main predictions, in particular
that every particle has a superpartner, that their spins
differ by $1/2$, that their gauge quantum numbers are the same, that
their couplings are identical, that certain mass relations hold, etc.
This will require 
precise measurements of masses, branching ratios,
cross sections, angular distributions, etc. A precise knowledge of as
many SUSY parameters as possible will be necessary to disentangle the
underlying pattern of SUSY breaking. In order to carry out this physics
program, experimental information from both the LHC and the LC will be
crucial.

Detailed studies of the possible interplay between LHC and LC in SUSY 
searches have been performed within the LHC / LC Study Group. It has been 
demonstrated that
experimental information on properties of uncoloured SUSY particles from
the LC can significantly improve the analysis of cascade decays at the
LHC~\cite{lhclcrep}. In particular, the precise measurement
of the LSP mass at the LC eliminates a
large source of uncertainty in the LHC analyses, improving thus the 
accuracy of the reconstructed masses of the particles in the decay
chain~\cite{lhclcrep}.

\begin{figure}
\resizebox{0.23\textwidth}{!}{%
  \includegraphics{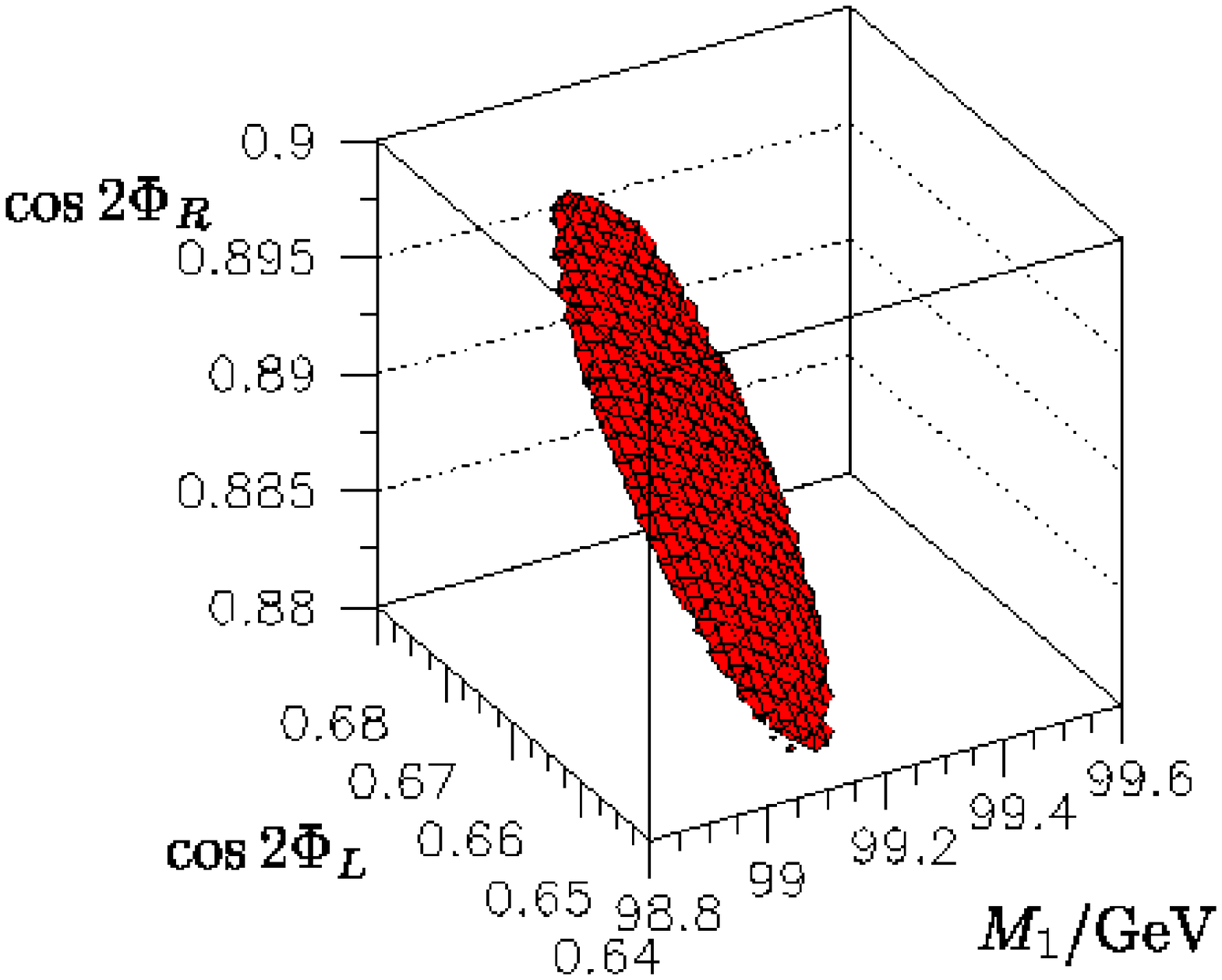}
}
\resizebox{0.23\textwidth}{!}{%
  \includegraphics{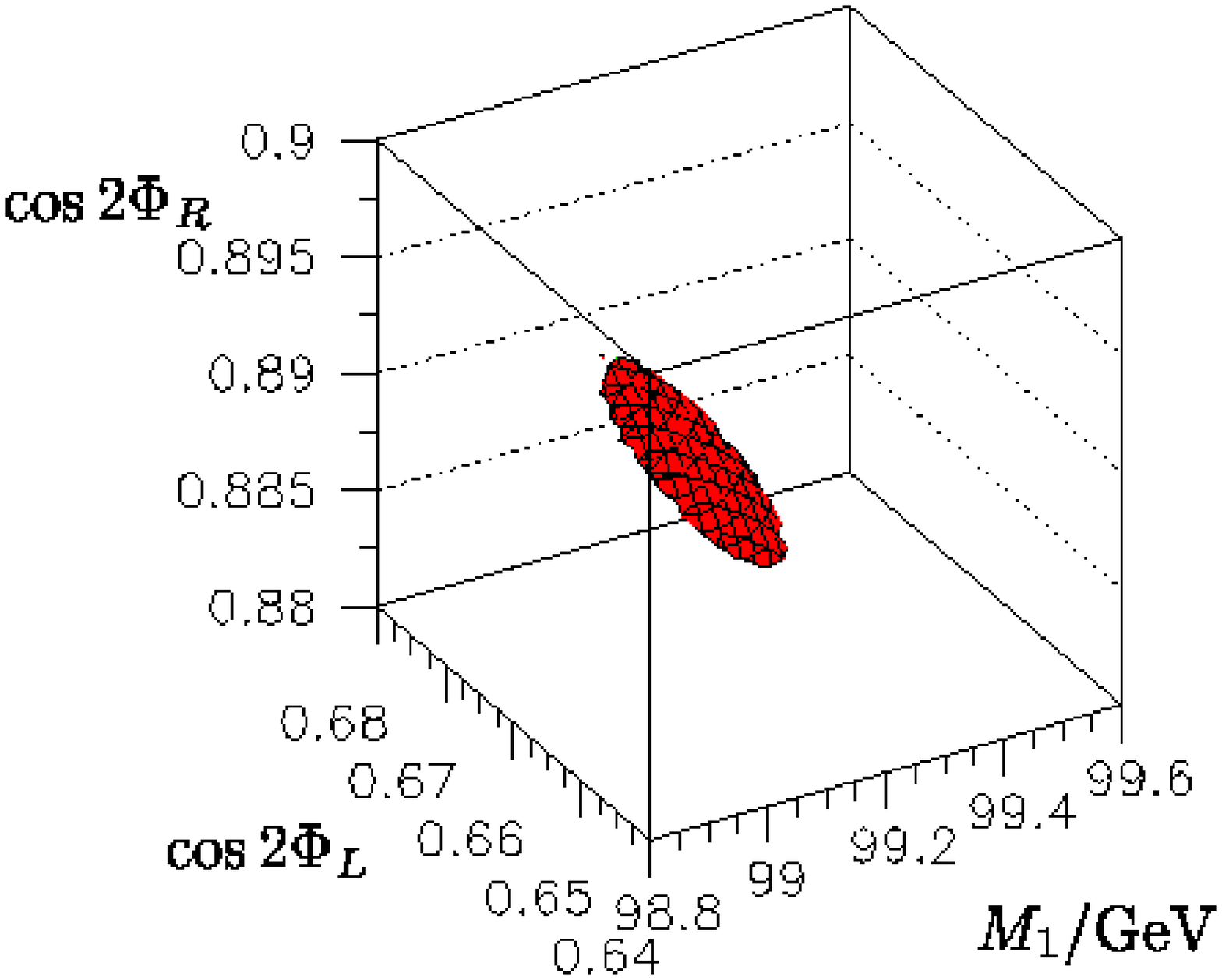}
}
\caption{
The $\Delta \chi^2=1$ contour in the  $M_1,\cos 2\phi_L,\cos 2\phi_R$
parameter space
derived from LC data alone (left) and from the joint analysis of the 
LC and LHC data (right), from \citere{maus}.
}
\label{fig:maus}
\end{figure}

In \citere{maus} an analysis was carried out based on the SPS~1a
benchmark scenario~\cite{sps} where the
measurement of the masses of the two lightest neutralinos, the lighter
chargino, the selectrons and the sneutrino at the LC was used to predict
the properties of the heavier neutralinos. It was demonstrated that this
input makes it possible to identify the heaviest neutralino at the LHC
and to measure its mass with high precision. Feeding this information back
into the LC analysis improves the determination of the fundamental SUSY 
parameters from the neutralino and chargino sector at the LC. In
\reffi{fig:maus} the accuracy in the determination of the parameter
$M_1$ (neutralino sector) and the mixing angles $\cos 2\phi_L$,
$\cos 2\phi_R$ (chargino sector) from LC data alone is compared with the
LHC / LC combined analyses, showing a significant improvement. This
results in a precise determination of the related parameters 
$M_1$, $M_2$, $\mu$ and $\tan\be$ and allows powerful consistency tests
of SUSY~\cite{maus}.

LC input on uncoloured SUSY particle properties together with
reconstructed sbottom masses from cascade decays can furthermore be used
to extend the LHC capabilities in the analysis of the stop and sbottom
sector~\cite{stoplhc}, leading to a reconstruction of the stop masses and
the stop and sbottom mixing angles~\cite{lhclcrep}. 



\subsection{SUSY Higgs physics}

The interplay between LHC and LC will also be crucial for exploring the
physics of electroweak symmetry breaking, both in its realisation via
the Higgs mechanism and via strong electroweak symmetry breaking. Many
detailed investigations can be found in \citere{lhclcrep}.

The LHC will discover a SM-like Higgs over the whole mass range 
$\mh \lsim 1$~TeV~\cite{lhcexp}. If the $H \to \ga\ga$ decay mode will be 
accessible, the LHC will be able to perform a first precision
measurement in the Higgs sector by determining the Higgs-boson mass with
an accuracy of about 
$\De\mh^{\rm exp} \approx 200$~MeV~\cite{lhcexp}. In contrast
to the SM, where $\mh$ is a free parameter, in SUSY models the mass of the 
lightest $\cp$-even Higgs-boson mass can be directly predicted from the
other parameters of the model. As a consequence of large radiative
corrections from the top and scalar top sector of the theory, the
prediction for $\mh$ sensitively depends on the input value of the
top-quark mass. The experimental error on $\mt$, which at the LHC will
be $\sim$1--2~GeV~\cite{lhcexp}, translates roughly
linearly into an uncertainty of $\mh$,
$\De\mh^{\de\mt} \approx \de\mt^{\rm exp}$~\cite{tbexcl}.
In order to match the experimental precision at the LHC, 
$\De\mh^{\rm exp} \approx 200$~MeV, with the accuracy of the theoretical
prediction, the precise measurement of the top-quark mass at the
LC, $\de\mt^{\rm exp} \lsim 100$~MeV~\cite{lctdrs}, will be mandatory. 
A precise measurement of $\mh$ and other Higgs sector observables
when compared with the MSSM prediction will allow to obtain
sensitive constraints on the MSSM parameters. 
In order to obtain indirect bounds on SUSY parameters in this way,
combined LHC / LC information on the other SUSY
parameters and a small residual uncertainty from unknown higher-order
corrections are crucial~\cite{delmt}.


\subsection{Higgs and radion searches at LHC and LC}

Models with 3-branes in extra dimensions typically imply the existence
of a radion, $\phi$, which can mix with the Higgs boson, thereby
modifying the Higgs properties and the prospects for its detectability
at the LHC. As a consequence, it might not be possible to observe a Higgs
boson at the LHC over a significant region of the parameter space
given by the Higgs-boson mass, $M_{\rm h}$, the radion mass, $M_{\phi}$,
the scale $\Lambda_{\phi}$, and the Higgs--radion mixing parameter,
$\xi$~\cite{radion}. For most of this parameter region the radion will
be observable in the process $gg \to \phi \to ZZ^* \to 4
\ell$~\cite{radion}, leading thus to a situation where one scalar will
be detected at the LHC. Disentangling the nature of this scalar state will
be a very important but experimentally challenging task.

\begin{figure}
\resizebox{0.23\textwidth}{!}{%
  \includegraphics{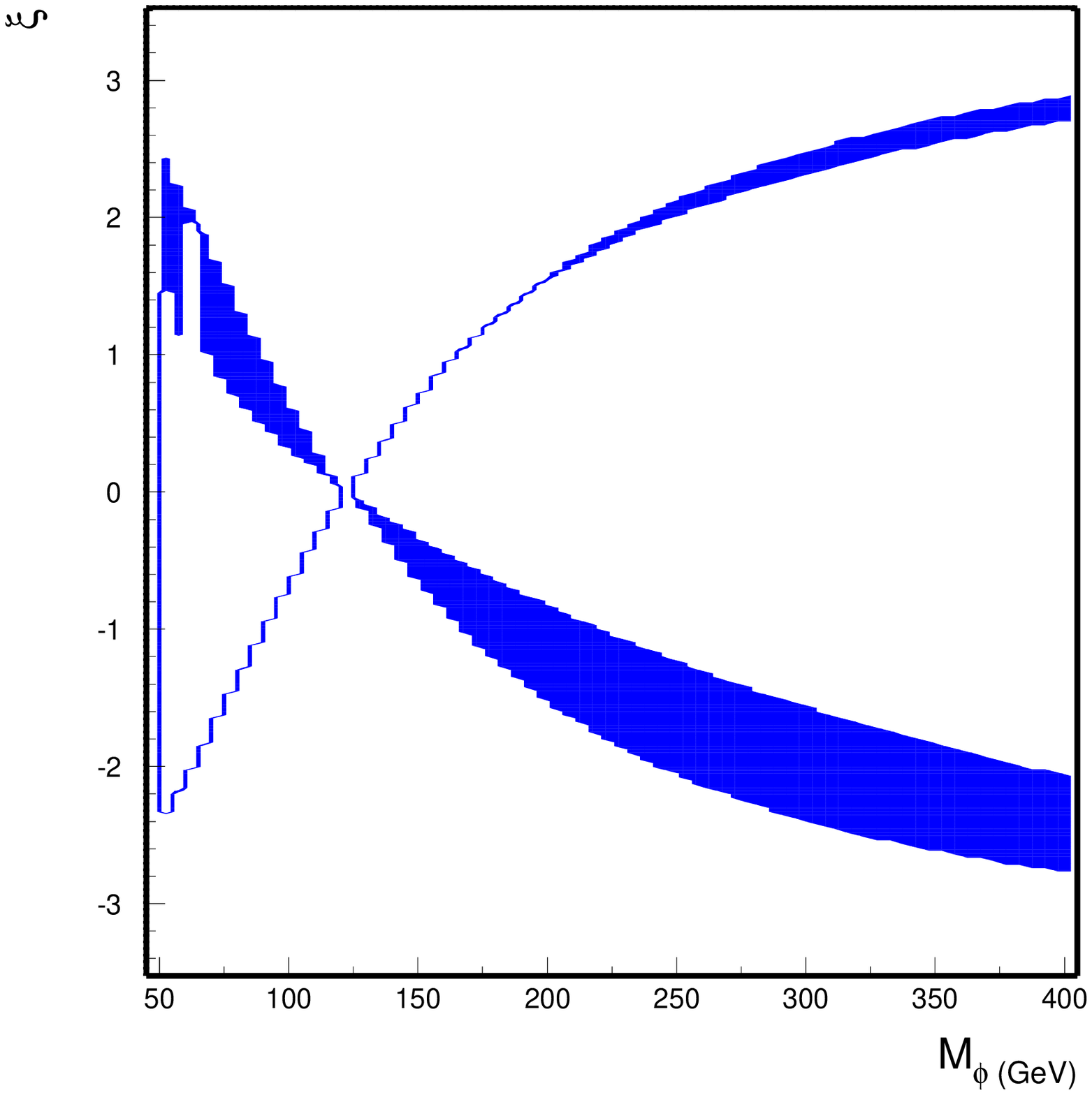}
}
\resizebox{0.23\textwidth}{!}{%
  \includegraphics{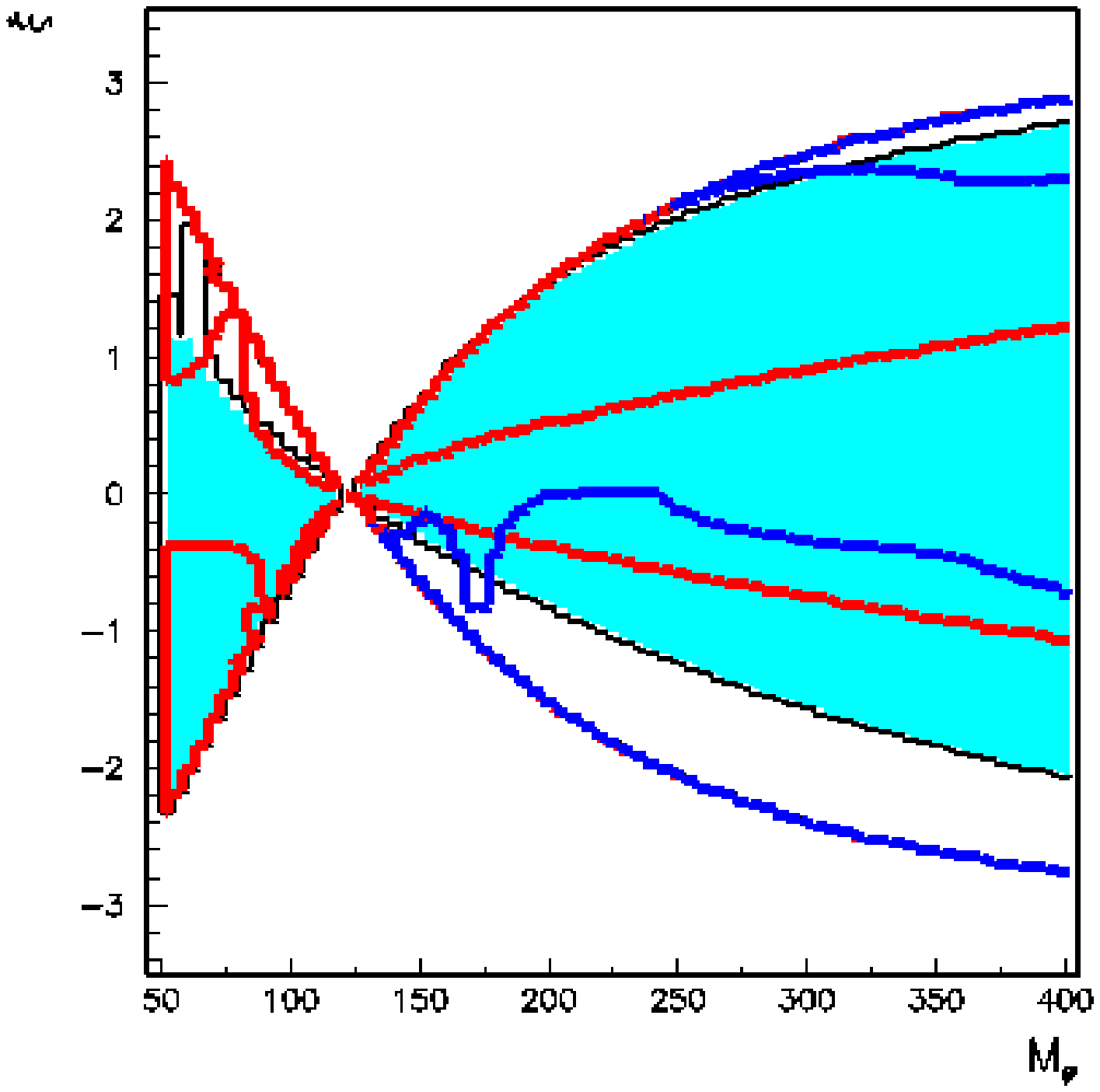}
}
\caption{
Parameter regions where the Higgs significance is below $5 \si$ at the
LHC for one experiment and 30~fb$^{-1}$ (left), regions (indicated by
the grey (red) lines, which extend also along the edges of the hourglass
region) where the precise measurements
of the $hb \bar b$ and $hWW$
couplings at the LC provide $> 2.5 \si$ evidence for the radion mixing
effect (right), from \citere{radion}.
}
\label{fig:radion}
\end{figure}

The LC should guarantee observation of the Higgs boson over the whole
parameter region and in addition observation of the radion even in most of the
regions within which detection of either at the LHC will be difficult.
Furthermore, precision measurements of the Higgs couplings to various
types of particle pairs will allow to experimentally establish the
Higgs--radion mixing effects. It is demonstrated in
\reffi{fig:radion} that the parameter regions for which the Higgs
significance is below $5 \si$ at the LHC (for $M_{\rm h} = 120$~GeV,
$\Lambda_{\phi} = 5$~TeV) overlap with the regions where precision
measurements of Higgs couplings at the LC establish the Higgs--radion
mixing effect~\cite{radion}. The LHC, on the other hand,
will observe the distinctive signature of 
Kaluza--Klein graviton excitation production over a substantial range of 
$\Lambda_{\phi}$ in these scenarios.

                                                                                
\subsection{New gauge theories and 
extra dimensions}

Many kinds of extensions of the SM lead to an enlarged gauge-boson
sector. Determining the nature of the new gauge bosons 
will require a variety of detailed experimental results
which can be provided by the interplay of LHC and LC. The LHC has a
large mass reach for direct detection of new gauge bosons, while the LC
has a large indirect reach arising from virtual effects of the new
states which result in deviations from the SM predictions. 

The LC, running at high energy, is sensitive to Z--Z$'$ interference effects 
through the fermion pair-production process, $e^+e^- \to f \bar f$,
yielding in particular the ratio of the Z$'f\bar f$ couplings and the 
Z$'$ mass.
If $M_{{\rm Z}'}$ is known from the LHC, the combined LHC / LC analysis
yields a determination of the Z$'$ couplings with high precision~\cite{zprime1}.
Furthermore, the measurements of the electroweak precision
observables in the GigaZ mode of the LC, i.e.\ $\sweff$, the total Z~width, 
the Z~partial widths and the W-boson mass, yield important information
for distinguishing different models of new physics~\cite{frichard}. This
input can be helpful for optimising the search strategies at the LHC.

Careful analyses are required in particular to distinguish a Z$'$ from
the lightest Kaluza--Klein excitations of the SM electroweak gauge
bosons~\cite{zprimeKK} and to determine the structure of Little Higgs
models from the properties of the new particle states~\cite{LittleHiggs}. 
Also in these cases it has been demonstrated that combined information
from LHC and LC can be crucial.



\bigskip

\noindent
{\bf Acknowledgements}\\
The author thanks the members of the LHC / LC Study Group 
for their fruitful collaboration.



\end{document}

%% file: defs.tex
\def\beq{\begin{equation}}
\def\eeq{\end{equation}}
\def\beqar{\begin{eqnarray}}
\def\eeqar{\end{eqnarray}}
\def\barr#1{\begin{array}{#1}}
\def\earr{\end{array}}
\def\bfi{\begin{figure}}
\def\efi{\end{figure}}
\def\btab{\begin{table}}
\def\etab{\end{table}}
\def\bce{\begin{center}}
\def\ece{\end{center}}

\def\text{\textstyle}

\def\be{\beta}
\def\ga{\gamma}
\def\de{\delta}

\def\si{\sigma}

\def\De{\Delta}

\def\reffi#1{\mbox{Fig.~\ref{#1}}}

\def\citere#1{\mbox{Ref.~\cite{#1}}}

\def\mathswitchr#1{\relax\ifmmode{\mathrm{#1}}\else$\mathrm{#1}$\fi}

\newcommand{\Ph}{\mathswitchr h}

\newcommand{\Pt}{\mathswitchr t}

\def\mathswitch#1{\relax\ifmmode#1\else$#1$\fi}

\newcommand{\Mt}{\mathswitch {m_\Pt}}

\newcommand{\mh}{\mathswitch {m_\Ph}}

\newcommand{\sweff}{\sin^2 \theta_{\mathrm{eff}}}

\newcommand{\mt}{\Mt}

\newcommand{\tsf}{\theta\kern-.20em_{\tilde{f}}}
\newcommand{\tsfp}{\theta\kern-.20em_{\tilde{f}\prime}}
\newcommand{\tsq}{\theta\kern-.15em_{\tilde{q}}}

\newcommand{\lsim}
{\;\raisebox{-.3em}{$\stackrel{\displaystyle <}{\sim}$}\;}

\newcommand{\cp}{{\cal CP}}

\newcommand{\VL}{\left( \begin{array}{c}}
\newcommand{\VR}{\end{array} \right)}
\newcommand{\ML}{\left( \begin{array}{cc}}
\newcommand{\MLd}{\left( \begin{array}{ccc}}
\newcommand{\MLv}{\left( \begin{array}{cccc}}
\newcommand{\MR}{\end{array} \right)}

\newcommand{\BC}{\begin{center}}
\newcommand{\EC}{\end{center}}
\newcommand{\BE}{\begin{equation}}
\newcommand{\EE}{\end{equation}}
\newcommand{\BEA}{\begin{eqnarray}}
\newcommand{\BEAnn}{\begin{eqnarray*}}
\newcommand{\EEA}{\end{eqnarray}}
\newcommand{\EEAnn}{\end{eqnarray*}}

\newcommand{\id}{{\rm 1\kern-.12em
\rule{0.3pt}{1.5ex}\raisebox{0.0ex}{\rule{0.1em}{0.3pt}}}}

\def\draftdate{\relax}
\def\mda{\relax}
\def\mua{\relax}
\def\mla{\relax}
\def\draft{
\def\thtystars{******************************}
\def\sixtystars{\thtystars\thtystars}
\typeout{}
\typeout{\sixtystars**}
\typeout{* Draft mode!
         For final version remove \protect\draft\space in source file
*}
\typeout{\sixtystars**}
\typeout{}
\def\draftdate{\today}
\def\mua{\marginpar[\boldmath\hfil$\uparrow$]%
                   {\boldmath$\uparrow$\hfil}%
                    \typeout{marginpar: $\uparrow$}\ignorespaces}
\def\mda{\marginpar[\boldmath\hfil$\downarrow$]%
                   {\boldmath$\downarrow$\hfil}%
                    \typeout{marginpar: $\downarrow$}\ignorespaces}
\def\mla{\marginpar[\boldmath\hfil$\rightarrow$]%
                   {\boldmath$\leftarrow $\hfil}%
                    \typeout{marginpar:
$\leftrightarrow$}\ignorespaces}
\def\Mua{\marginpar[\boldmath\hfil$\Uparrow$]%
                   {\boldmath$\Uparrow$\hfil}%
                    \typeout{marginpar: $\Uparrow$}\ignorespaces}
\def\Mda{\marginpar[\boldmath\hfil$\Downarrow$]%
                   {\boldmath$\Downarrow$\hfil}%
                    \typeout{marginpar: $\Downarrow$}\ignorespaces}
\def\Mla{\marginpar[\boldmath\hfil$\Rightarrow$]%
                   {\boldmath$\Leftarrow $\hfil}%
                    \typeout{marginpar:
$\Leftrightarrow$}\ignorespaces}
\overfullrule 5pt
\oddsidemargin -15mm
\marginparwidth 29mm
}